\documentclass{ws-ijmpd}
\usepackage{graphicx}

\def\arcsec{\hbox{$^{\prime\prime}$}}

\begin{document}

\markboth{D. A. Schwartz}
{\emph{Chandra} X-ray Observatory}

\catchline{}{}{}{}{}

\title{THE DEVELOPMENT AND SCIENTIFIC IMPACT OF THE \emph{CHANDRA}
  X-RAY OBSERVATORY} 

\author{Daniel A. Schwartz}
\address{Harvard/Smithsonian Center for Astrophysics, 60 Garden St.,\\
  Cambridge, MA, 02138, USA \\ das@head.cfa.harvard.edu}


\maketitle

\begin{history}
\received{(DAY MONTH YEAR)}
\revised{(DAY MONTH YEAR)}
\end{history}

\begin{abstract}
I review the operational capabilities of the \emph{Chandra} X-ray
Observatory, including some of the spectacular results obtained by the
general observer community.  A natural theme of this talk is that
\emph{Chandra} is revealing outflows of great quantities of energy
that were not previously observable. I highlight the \emph{Chandra}
studies of powerful X-ray jets.  This subject is only possible due to the
sub-arcsecond resolution of the X-ray telescope. 
\end{abstract}

\keywords{X-ray astrophysics; Supernova nebulae; Cluster cooling
  flows; Quasars: jets}


\section{Development of the \emph{Chandra} X-ray Observatory}

The Advanced X-ray Astrophysics Facility (\emph{AXAF}) was developed
to serve the entire astronomical community for attacking a wide range
of problems.\cite{weisskopf87,weisskopf96,weisskopf00} The
requirements to do this included large telescope area, access to the
entire sky with  more than 85\% available at any time, high observing
efficiency and a long operational lifetime, instruments which could
provide imaging and spectroscopy, including spatially resolved
spectroscopy with at least modest (E/$\Delta E \approx$ 10 to 50)
energy resolution, and the ability to locate the measured photons on
the sky. However, by far the most stringent and crucial requirement
was for the telescope to be able to image to better than 0.5\arcsec\
FWHM. More precisely, it was required that a 1\arcsec\ diameter circle
about a point source would contain 70\%, and 20\%, of the photons
imaged at 1 and at 8 keV, respectively. This allows high contrast for
large dynamic range, and gives an imaging point spread function (PSF)
which is not a strong function of energy. Prior to launch, \emph{AXAF}
was renamed the \emph{Chandra} X-ray Observatory (CXO), in honor of
the Indian-American astrophysicist Subramanyan Chandrasekhar.

The exquisite X-ray imaging capability of the High Resolution Mirror
Assembly (HRMA) is the key to the scientific power of the
\emph{Chandra} observatory. Imaging to 0.5\arcsec\, in contrast to the
$\sim$5\arcsec\ imaging capability of the previous \emph{Einstein} and
\emph{ROSAT} X-ray missions, is truly a 100-fold improvement in the
imaging capability. This enables \emph{Chandra} to study jets and
outflows in quasars and active galaxies which had previously been
considered ``point'' sources, to reveal structure and interactions
within clusters of galaxies for which the distribution of hot gas had
previously been considered smooth and symmetric, and to allow point
source detection to fluxes 100 times fainter due to the reduction of
the detector area which accumulates background.

One of two different imaging instruments can be moved into the
telescope aim point at any time. The High Resolution Camera
(HRC)  uses micro channel plates (MCP) to convert the X-ray, and to amplify
the resulting electrons. The X-ray is located by determining the
centroid of the resulting electron cloud. The Advanced Camera for
Imaging Spectroscopy (ACIS, formerly the ``AXAF Camera for Imaging
Spectroscopy''), uses charge coupled devices (CCD) to convert X-rays into a
number of electrons directly proportional to the photon energy, and
clocks these out in a fixed pattern which indicates the position at
which the X-ray was imaged. A low energy transmission grating (LETG)
gives dispersive resolution optimized for the 0.07 to 2 keV range,
while a high energy transmission grating (HETG) consists of two
separate sets of  gratings optimized for the 0.4 to 5 keV range (medium
energy grating) and the 0.8 to 10 keV range (high energy grating). If
desired, either the LETG or HETG (but not both!) can be inserted into
the optical path.

A unique feature of celestial X-ray astronomy is that it is based on
single photon counting.  The energies are large enough, from 0.1 to 10
keV in the case of \emph{Chandra}, that this is possible, and from
most celestial sources the fluxes are so low that it is imperative.
At the limiting sensitivity in the deepest fields, 10 photons in a two
week integration represents a significant detection.  Operationally
this means that the satellite need not point rigidly to within a very
small fraction of the desired resolution for the duration of the
exposure.  In fact, it is desireable to force the telescope to dither
over many pixels in order to use an averaged calibration, since it is
not feasible to calibrate each resolution element to the accuracy
ultimately desired. An optical CCD camera which simultaneously
measures stars on the sky and fiducial lights on the instruments
enables post facto ground reconstruction of the image to better than
0.1\arcsec\ .

\section{\emph{Chandra} Studies of Energy Outflows}

X-ray astronomy, and \emph{Chandra} in particular, studies virtually
every type of astronomical system.  To limit the scope of this review
I have chosen some examples where the \emph{Chandra} observations are
giving us information on the outflow of matter and energy. This
presents a contrast to much of the historical impact of X-ray
astronomy, which has been to elucidate accretion processes on compact,
binary stars in our galaxy, and on supermassive black holes in active
galactic nuclei (AGN). Specifically, I will show some results on
supernovae and pulsar wind nebulae, on X-ray ``cavities'' found in
clusters of galaxies and the fate of cooling flows of gas in those
clusters, and on X-ray jets seen in quasars and powerful FR II radio
galaxies and the implication that these may serve as ``cosmological
beacons.''

\subsection{Pulsar wind nebulae}

Gaensler\cite{gaensler03} has recently reviewed the winds expelled
from pulsars, bringing to bear multifrequency radio, optical, and
X-ray observations. From the early radio observation of pulsars, and
their interpretation as neutron stars, the rate of period increases
led to a realization that up to 10$^{39}$ ergs s$^{-1}$ of rotational
energy was being lost. This dwarfed the amount of energy detected as
electromagnetic emission across the radio to $\gamma$-ray
spectrum. Theory suggested that the bulk of the power went into 
acceleration of relativistic particles\cite{michel69a} and generation
of a wind.\cite{michel69b} The wind would emanate from the speed of
light circle at radius $r=c/f$, where $c$ is the speed of light and
$f$ the pulsar frequency.  This region cannot be imaged, being much
smaller than microarcseconds. 

\begin{figure}[h]
\includegraphics[viewport= 105 50  545 760,clip,angle=-90,width=4.in]{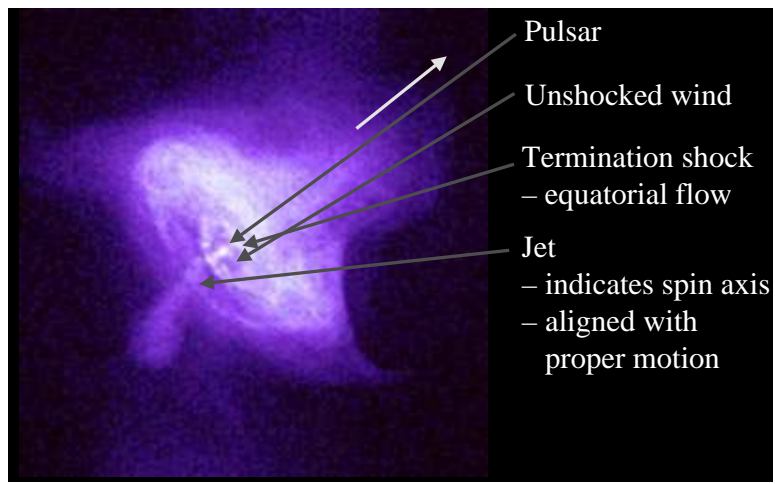} 
\caption{ \label{crabfig} \emph{Chandra} image of the Crab Nebula
(NASA/CXC/MSFC from Ref. 4), showing the pulsar, the
termination shock from the relativistic equatorial wind, and the jet.
The large arrow shows the direction of the spin axis and the proper
motion.}
\end{figure}

What has now been observed, as shown in Figure \ref{crabfig} for the
Crab Nebula, is the
termination shock where the relativistic wind hits the nebula ejected
by the supernova. The apparently concentric rings indicate an
outflow, which can be assumed as equatorial by symmetry. The
jet which appears perpendicular to the rings must be associated with
the rotational axis, as the magnetic axis must sweep a wide region on
the sky in order to modulate the pulses.

\begin{figure}[h]
\resizebox*{4.999in}{2.3in}{\includegraphics[viewport= 380 -60  615 380,clip,angle=-90]{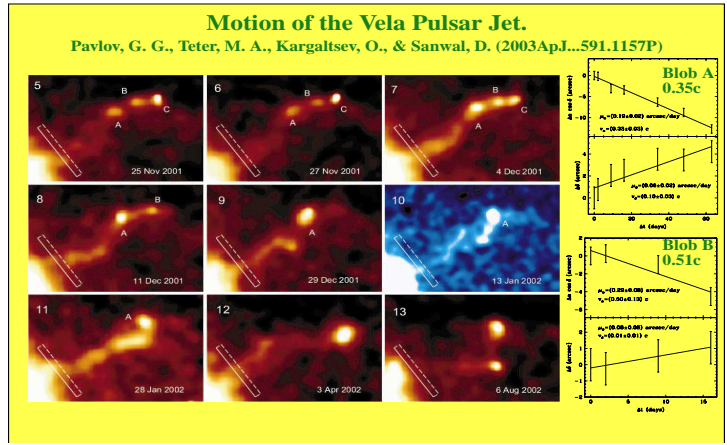} }
\caption{\label{velajet} Sequence of \emph{Chandra} images of the
  outer jet of the Vela pulsar (from Ref. 9) showing changes in morphology, and systematic outward motions.}
\end{figure}

The structure of the Crab Nebula wind and jet is known to have
remarkable optical and X-ray variations, and apparent motions at half
the speed of light.\cite{hester02} The Vela Pulsar Nebular also shows
a remarkable time sequence of changes in its outer jet
(Figure \ref{velajet}).\cite{pavlov03} From this one can infer apparent
relativistic motion\cite{pavlov03} of 0.35$c$ and 0.51$c$ for the
blobs A and B, respectively, in Figure \ref{velajet}.

\subsection{Cooling flows in clusters of galaxies}

A major problem on which \emph{Chandra} has provided key data is that
of cooling flows in clusters of galaxies (e.g.,
Ref.~\refcite{fabian94}).  Hot gas of temperature $\sim 10^8 \,^\circ
K$ fills rich clusters of galaxies, and emits X-rays via thermal
bremsstrahlung.  From the X-ray surface brightness, one can deduce the
electron density profile. Typically, in the core of the cluster the
density is sufficiently high that the cooling time becomes much less
than the Hubble time. This implies that the cooled gas should fall
into the cluster core, becoming more dense and cooling at an ever
increasing rate.  Indeed, the X-ray brightness profile of such
clusters shows emission highly peaked at the cluster center. Infall
rates in excess of 100 solar masses per year were often
deduced.\cite{arnaud88}  However, the destination of the gas was a
mystery. Star formation rates were no more than  10 solar masses
per year. Spectroscopic observations did not reveal the line emission
expected from X-ray emitting gas at temperatures below 1 to 2 keV.

It now seems that ``cavities,'' depressions seen in
the X-ray surface brightness, are indicators of energy input from the
central radio source.  These cavities, also called ``holes'' or
``bubbles,'' were first detected in \emph{ROSAT} observations of the
Perseus Cluster,\cite{bohringer93} and are detected by \emph{Chandra}
in at least 18 cluster atmospheres.\cite{birzan03}  

\begin{figure}
\includegraphics[viewport= 212 25  632 596,clip,angle=-90,width=4.in]{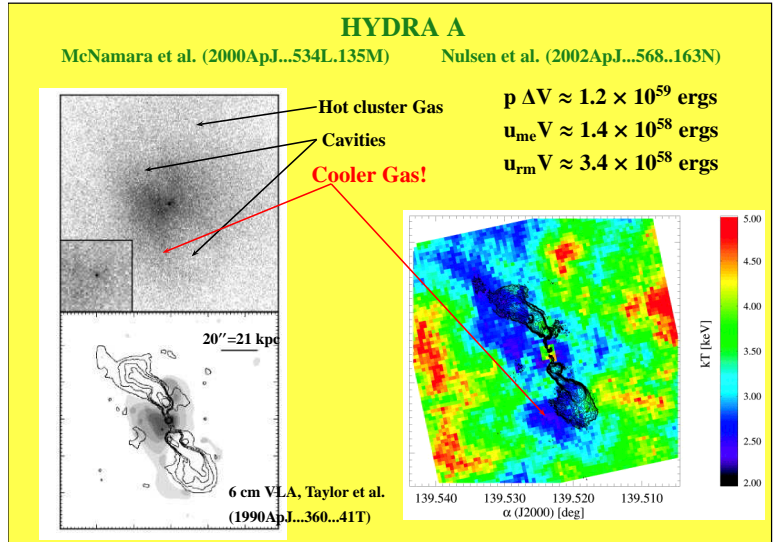} 
\caption{\label{hydraA} X-ray and radio images of the center of the Hydra A cluster of galaxies.}
\end{figure}

Figure \ref{hydraA}, from Refs.~\refcite{mcnamara00} and
\refcite{nulsen02}, shows studies of the Hydra A cluster. The
association of cooling flows with central powerful radio sources had
been documented,\cite{burns90} and it had been suggested that the
energy from radio sources in a central cD galaxy could balance the
radiative losses and stop the cooling flows, (e.g.,
Ref.\refcite{tucker83}). In this picture, it was assumed that a radio
jet would cause shock heating upon colliding with the cluster
gas.\cite{clarke97,heinz98} Consistent with this prediction, the lower
left panel in Figure \ref{hydraA} shows a two-sided, 5 GHz
jet\cite{taylor90} from the central galaxy entering the two main
cavities and filling them with radio plasma.  However, contrary to the
predictions, the higher density gas around the cavity rims is at a
\emph{lower} temperature than the ambient cluster gas, as shown in the
lower right hand panel\cite{nulsen02} in Figure \ref{hydraA}. 

The interpretation is that these cavities are bubbles inflated by work
done by the radio jet, and which are now rising buoyantly through the
cluster. Since the volume is directly observed, assuming rotational
symmetry, and the gas density and temperature are deduced from the
X-ray surface brightness and spectrum, one can calculate the work done
to expand the cavity. For the northern Hydra A cavity, p$\Delta \rm
{V} = \rm {kT} \Delta \rm {V} \approx$1.2 $\times$ 10$^{59}$
ergs. However, if one deduces the minimum energy density in particles
and magnetic field which gives the observed radio emission, only about
1.4 $\times$ 10$^{58}$ ergs is available. In the case of the Perseus
Cluster, the discrepancy is almost two orders of magnitude between the
work p$\Delta \rm {V} \approx$2 $\times$ 10$^{58}$ to evacuate the
North cavity and the available minimum energy.  These are not
necessarily discrepancies.  The radio plasma may deviate from the many
assumptions made to carry out the equipartition calculation. In the
case of Hydra A, Ref.~\refcite{taylor90} argue that the rotation
measure implies a magnetic field several times less than
equipartition, giving a total energy several times larger.  Also, the
above calculations all ignored the bulk kinetic flux which might have
been carried by a relativistic jet.  In the next section, we will see
that bulk relativistic motion in the jet is probably a general feature
of quasars and FR II radio sources, and may carry a kinetic energy
flux exceeding the radiative luminosity.

\subsection{X-ray jets in quasars}

Prior to \emph{Chandra}, extragalactic X-ray jets had been detected in
the famous objects Cen A, M87, and 3C 273. The sub-arcsecond imaging
capability of \emph{Chandra} has opened up the general study of X-ray
jets in radio sources.  Although jets may extend 5\arcsec\ to
20\arcsec\ from the quasar or radio galaxy, their surface brightness is lost
in the core PSF when the latter is larger than a few arcseconds,
because the jet flux is typically only one to a few percent that of the 
core.  In general, for the low power, FR I type
radio sources the X-ray jet emission can be interpreted as an
extension of the radio synchrotron emission.\cite{worrall01} Here I
will discuss only the powerful, FR II galaxies and quasars.

\emph{Chandra} has made many serendipitous observations of jets,
e.g. in PKS 0637-752,\cite{schwartz00} PKS~1127-143,\cite{siemiginowska02}  
and GB 1508+5714,\cite{siemiginowska03} and has pointed at some of the well
known objects, notably 3C 273.\cite{marshall01}  Two
groups\cite{sambruna02,marshall02} have attempted systematic X-ray
studies of FR II type radio jets.  I will discuss four objects from
the latter survey to illustrate the conditions which are generally
being derived for these systems.\cite{schwartz03}  

\begin{figure}
\includegraphics[viewport= 28 25  632 760,clip,angle=-90,width=3.7in]{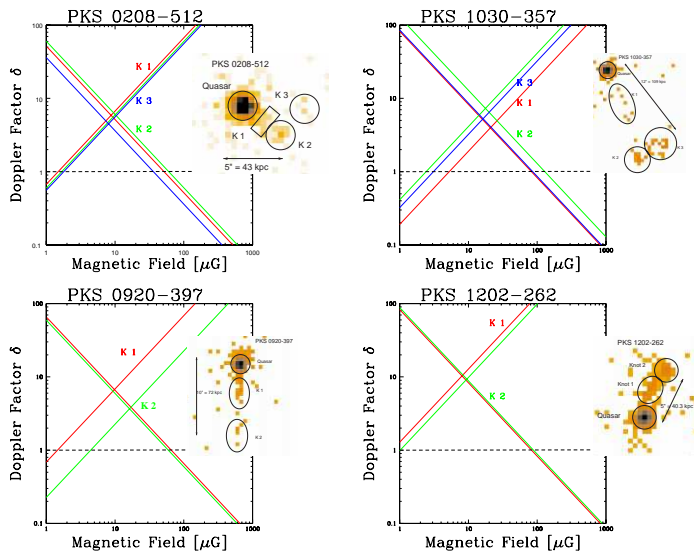} 
\caption{\label{jetdata} \emph{Chandra} data from four quasars with
  radio jets. Intersections of the lines of equipartition,
  $\delta \propto 1/B$, with the lines of IC/CMB production of X-rays,
  $\delta \propto B$, give solutions for B and $\delta$ for each region.}
\end{figure}

These four objects are shown in Figure \ref{jetdata}. For each of the
regions indicated we compare the fluxes we have measured at 8.6 GHz,
and in some cases also at 4.8 GHz. For most regions our optical upper
limits\cite{gelbord03} prohibit extrapolating the radio flux into the
X-ray region. This shows that the X-rays are not synchrotron radiation
from the same population of electrons which are giving rise to the
radio emission.  However, the close spatial
correlation of X-ray and radio emission indicates that they do arise
from the same sources. This leaves inverse Compton (IC) scattering of the
electrons as the most likely mechanism for X-ray emission. IC
against the cosmic microwave background (CMB) is an inevitable process
if the jet is in bulk relativistic motion relative to the CMB with
Lorentz factor $\Gamma$ greater than a few units.\cite{tavecchio00,celotti01} 

When we interpret all the jet regions in Figure \ref{jetdata} as IC/CMB
emission we derive Doppler factors $\delta = (\Gamma(1-\beta \cos
\theta))^{-1}$ between 3 and 12, and equipartition magnetic fields in
the jet rest frame between 5 and 25 $\mu$G. From these numbers we can
calculate the kinetic flux through each element following
Ref.~\refcite{ghisellini01}, as K=A$\Gamma^2\,c\,$U, where A is the
measured cross sectional area (assuming cylindrical symmetry) and U =
U$_{B}$ +U$_{e}$ +U$_{p}$ is the energy density, written as the sum of
the magnetic, electron, and proton energy densities.  We assume
equipartition, U$_{B}$ =U$_{e}$, and parameterize U$_{p}$=k
U$_{e}$ and take k=1. We find kinetic
fluxes of order 10$^{46}$ to 10$^{47}$ ergs s$^{-1}$ for these
jets. These numbers are comparable to or larger than the bolometric
radiation from the quasar, in line with theoretical considerations
showing that jets are an enegetically important component of accretion
processes.\cite{meier03} We also see that the jets transport energy with large efficiency. The ratio of the emission
radiated in the jet frame to the kinetic flux is of order
10$^{-5}$.

\section{Conclusions}

\emph{Chandra} observations have discovered the evidence of large
outflows of energy and material previously  not seen directly. In the
case of pulsar wind nebulae observations prove that jets are ejected
along the spin axis, and that material is also ejected in the
equatorial plane. Theoretically these would be oppositely charged
plasmas.\cite{michel00} In clusters of galaxies, the images prove a close
X-ray radio connection. In quasars we have seen that the kinetic fluxes
carried by jets represent very large energy contents. We might
hypothesize that jets from the radio galaxies at the center of cooling
flow clusters may carry similarly large energy fluxes -- clearly
adequate to compensate for the radiative cooling of the gas, even if
we allow that the radio source only has a $\sim$10\% duty cycle. To
actually compensate for a cooling flow probably requires a feedback
mechanism involving the infalling gas and the central
AGN.\cite{nulsen03} Observations of the Perseus cluster of galaxies
led to the suggestion that sound waves provide the mechanism for
dissipation of the jet energy quasi-isotropically through the
cluster.\cite{fabian03} 

If the X-ray emission from jets does arise from the IC/CMB, then the
same objects which we have already seen would be easily detectable at
any arbitrarily larger redshift at which they might
occur.\cite{schwartz02} This is because the emission is proportional
to the energy density of the target photons, which will increase as
T$^4\propto$T$_{0}^4$ (1+z)$^4$, compensating exactly for the
cosmological diminution of surface brightness.

This review has been based primarily on the first three years of
\emph{Chandra} operation. We are just finishing the fourth year and
starting the approved fifth year program. At the \emph{Chandra} X-ray
Center we have performed a study indicating the expected lifetime of
the observatory will be at least 15 years. We expect these future
observations to unveil the physical processes in all astronomical
systems, including the few topics included in this review, in vastly
greater breadth and detail.


\section*{Acknowledgements}

This work was supported in part by NASA contract NAS8-39073 to the
Chandra X-ray Center, and NASA grants GO2-3151C and G03-4120X to
SAO.


\end{document}